\documentclass[12pt,preprint,numberedappendix,twocolappendix]{emulateapj}
\newcommand\bibinc{n}		% set to y if bib pasted in .tex, set to n to use bibtex

\usepackage{lineno}
\usepackage{subeqnarray}
\usepackage{amsmath}
\usepackage{hyperref}
\usepackage{ulem}
\usepackage{stmaryrd}

\hypersetup{colorlinks=true,linkcolor=red,citecolor=blue,urlcolor=blue}

\usepackage{natbib}
\def\tt{\bf   }

\def\mps{\;{\rm m~s^{-1}}}
\def\jpkgk{\;{\rm J~kg^{-1}~K^{-1}}}

\newcommand {\twostr} {{\ttfamily TWOSTR}}

\setcitestyle{citesep={,}}
\begin{document}

\slugcomment{Accepted to AAS Journals}

\shorttitle{Convection in pure-steam atmospheres}
\shortauthors{Tan,  Lef\`evre \& Pierrehumbert }

\title{Convection modeling of pure-steam atmospheres}
\author{Xianyu Tan, Maxence Lef\`evre and Raymond T. Pierrehumbert}
\affil{Atmospheric Oceanic  and Planetary Physics, Department of Physics, University of Oxford, OX1 3PU, United Kingdom}
\email{xianyu.tan@physics.ox.ac.uk}
\begin{abstract} Condensable species in either vapor or condensed form are crucial in shaping planetary climate. A wide range of  planetary climate systems involve understanding  non-dilute condensable substances and their influence on climate dynamics.  There has been progress on large-scale dynamical effects and on 1D convection parameterization, but resolved 3D moist convection remains unexplored in non-dilute conditions, though it can have a profound impact on temperature/humidity profiles and cloud structure.  In this work, we tackle this problem for pure-steam atmospheres using three-dimensional, high-resolution numerical simulations of  convection in post-runaway atmospheres where the water reservoir at the surface has been exhausted.  We show that the atmosphere is comprised of two characteristic regions, an upper condensing region   dominated by gravity waves and a lower noncondensing region characterized by convective overturning cells. Velocities in the condensing region are much smaller than those in the lower  noncondensing  region, and the  horizontal  temperature  variation  is small ($\lesssim1$ K) overall. Condensation  in the thermal photosphere  is largely driven by radiative cooling and tends to be statistically homogeneous. Some condensation also happens deeper, near the boundary of the condensing region, due to triggering by gravity waves and convective penetrations and  exhibit random patchiness. This qualitative structure is insensitive to varying parameters in the model, but quantitative details may differ. Our  results confirm theoretical expectations that atmospheres close to the pure-steam limit do not have organized deep convective plumes in the condensing region.  The generalized convective parameterization scheme discussed in \cite{ding2016} is appropriate to handle the basic structure of atmospheres near the pure-steam limit but is difficult to capture gravity waves and their mixing that appear in 3D convection-resolving models.  
\end{abstract}

\section{introduction}

Characterization of the smaller exoplanets is beginning to come into view,  (e.g., \citealp{kreidberg2014,knutson2014,benneke2019,tsiaras2019}), and their atmospheres can present
greater compositional diversity than the $\mathrm{H_2/He}$ dominated atmospheres of hot Jupiters. Condensable species, either in  vapor or condensed form, 
%could be one of the major constituents in planetary atmospheres and 
are important in determining planetary climate
%via their radiative and latent heat effects 
\citep{pierrehumbert2010}. 
The need to better understand the planetary climate dynamics with non-dilute condensable constituents is demanding. Non-diluteness refers to a situation wherein the mass of condensable substances can be comparable to that of the noncondensing components. Exoplanets with sizes in between Neptune and super-Earths  may have water-rich atmospheres (e.g., \citealp{zeng2019,mousis2020,otegi2020,harman2021}). Terrestrial planets inward of the habitable zone may experience a runaway greenhouse   wherein the ocean evaporates and water vapor dominates the atmosphere \citep{kasting1993}. { Steam atmospheres are relevant to terrestrial planets during the magma ocean phase immediately following accretion (e.g., \citealp{zahnle1988,hamano2013}), and for young planets whose water vapor has not yet condensed into an ocean \citep{turbet2021}. }   Extremely close-in rocky planets may have rock-vapor atmospheres on    the dayside which are  condensable during the transport to the nightside \citep{castan2011}.  Finally, Mars and (marginally) Titan are in this regime as well with the major condensable constituents being ${\rm CO_2}$ and ${\rm CH_4}$, respectively.

Several  studies have targeted the climate dynamics  with non-dilute condensable vapor. \cite{ding2016}  examined the energy budget associated with non-dilute condensation and precipitation, and  proposed a  scheme for convection parameterization applicable in general conditions.  \cite{pierrehumbert2016}   demonstrated the novelty on large-scale dynamics arisen from mass transport by precipitation and constraints from the condensation thermodynamics, then  presented general circulation models (GCMs) in non-dilute conditions. \cite{yamashita2016} performed two-dimensional convection modeling for a pure-${\rm CO_2}$ atmosphere in Martian conditions. \cite{ding2018} further demonstrated the importance of horizontal heat transport in determining global surface temperature variation of pure-steam atmospheres. 
%Using GCMs developed in \cite{pierrehumbert2016}, 
\cite{ding2020} showed that phase-curve information may be used to distinguish richness of water vapor in atmospheres of slowly-rotating, tidally-locked terrestrial planets. {\cite{turbet2021} performed long-term GCM simulations for early hot steam atmospheres of Earth and Venus and suggested the important role of cloud radiation effects.}

{ Understanding the nature of convection is vital as it is an important form of energy and mass transport in the atmospheres and can profoundly  impact climates.  Three dimensional resolved convection
simulations are a valuable tool for advancing the understanding of convection, and for testing the parameterizations that are essential for representing convection in most global circulation models. 
Convection modeling has been carried out in the context of exoplanets  \citep{zhang2017surface,sergeev2020,lefevre2021,song2021}, although  mostly in Earth-like conditions. Convection in non-dilute conditions is in a novel regime that is little explored, though work on 1D hydrostatic parameterizations provides some hypotheses as to the expected behavior \citep{ding2016}.   1D radiative-convective models of pure-steam atmospheres typically assume a temperature structure of a dry adiabat attached to the surface temperature and a dew-point adiabat of water vapor within the saturated region \citep{goldblatt2013,hamano2013,Boukrouche2021}. This assumption needs to be validated using convection models that self-consistently capture  relevant dynamical and condensation processes. }

In this work, we investigate the 3D nature of convection for the limiting case of a pure-steam atmosphere. {This is easier to understand than general nondilute conditions  and  sets a baseline for future work that incorporates varying fractions of non-condensable gases,
but there are also numerous planetary phenomena for which the pure-steam limit is relevant. }
In Section \ref{ch.model}, we introduce our numerical model; in Section \ref{ch.results} we present our basic results and sensitivity exploration; and finally we discuss and conclude in Section \ref{ch.discussion}. 

\section{Model}
\label{ch.model}
\subsection{CM1}
We utilize the Cloud Model 1 (CM1), a   3D, non-hydrostatic   model that has been widely used for  convection studies (\citealp{bryan2002}, see also \url{https://www2.mmm.ucar.edu/people/bryan/cm1/}), to perform high-resolution, idealized   experiments of convection over a regional domain.  We use a plane-parallel, two-stream radiative transfer scheme with a grey approximation in the thermal emission. The numerical tool \twostr ~\citep{kst1995} is employed to solve the radiative transfer equations and is coupled to the dynamics of CM1. The gas opacity of water vapor is assumed to be $0.1\;{\rm m^2kg^{-1}}$, same as that used in \cite{ding2016}. We apply a fixed surface temperature as a lower boundary condition  and treat it as a free parameter. Similar lower boundary conditions have been widely used in   modeling studies of convective aggregation (see a recent review by \citealp{wing2017}) as well as 1D radiative-convective models for runaway climate \citep{Boukrouche2021}. We assume a solid surface, and   there is no evaporative surface flux to the atmosphere. This is a post-runaway situation, after the ocean has completely evaporated into the atmosphere leaving a subsaturated layer near the ground. The atmosphere is assumed to be transparent to instellation. Note that water vapour has  strong near-IR absorption, which allows the incoming stellar radiation to heat the atmosphere.  In this paper, the effect is neglected so as to focus on the essentials of the problem.  We utilize the   fully-compressible equation set with only  water vapor in our models, but with a scheme to deal with condensation, rainout and evaporation as described below.
This study is the first extraterrestrial use of the CM1.

\subsection{Condensation, rainout and evaporation}
{ In \cite{ding2016} the First Law of Thermodynamics was used to adjust an entire atmospheric
column to an energetically consistent state neutrally stable to convection, after buoyancy-generated kinetic energy has been dissipated as heat.  Here, because we use a nonhydrostatic
model that explicitly resolves the conversion of potential to kinetic energy by buoyancy, the mixing it causes, its subsequent dissipation as heat,\footnote{{ In our 3D convection resolving model, the sources of kinetic energy dissipation include surface stress from the lower boundary layer, parameterized subgrid turbulent dissipation, and numerical dissipation. The dissipation of kinetic energy turns into a heat source in the thermodynamics equation, and this is a built-in option in the CM1 model. }} and the pressure adjustment triggered by the removal
of precipitation, our only use of the First Law is to determine the amount of condensate
produced or evaporated within individual grid cells, and the effects on local pressure and temperature.
{ An important difference between our 3D modeling and the 1D parameterization scheme \citep{ding2016} is that the adjustment of the column is handled explicitly by the dynamical core of our 3D model.  The adjustment occurring in our model -- which is not required to be complete -- is a natural emergent property of the dynamics.}

A two step process is modeled, in which precipitation is first produced {\it in situ} and then
removed by rainout. Let's suppose that we start with a mass of atmosphere with no condensate
present, but that radiative cooling or adiabatic ascent creates some supersaturation; a small
amount of condensate will form, the pressure will adjust to the phase boundary, and the latent
heat release will slightly increase the temperature of the parcel. 
Without energy and mass exchange with the environment during condensation, the air parcel follows the following relation based upon the first law of thermodynamics and mass conservation \citep{emanuel1994,ding2016}:  } 
%With an arbitrary mixing ratio of condensable constituents, the isolated atmospheric column conserves the following quantity \citep{ding2016}:
%\begin{equation}
%    \int_{z=0}^{\infty} \rho\left(k-\frac{p}{\rho}+gz\right)dz,
%    \label{eq.enthalpy}
%\end{equation}
\begin{equation}
    d\left( k-\frac{p}{\rho}\right) + pd\frac{1}{\rho} = 0,
    \label{eq.enthalpy}
\end{equation}
{where $\rho$ is the atmospheric density which includes all gases and the condensates, $k=c_{pl}q_t T+Lq_c$, $c_{pl}$ is that for the condensates, $L$ is the latent heat, $q_t$ is the mass concentration of the total condensable in both phases, and  $q_c$ is the mass concentration of the condensable vapor. $k$ is the  moist enthalpy in the pure steam limit, in which we also
have $q_t = 1$. Because CM1 is a nonhydrostatic model using $z$ rather than $p$ as vertical
coordinate, we have written the First Law in constant-volume rather than constant-pressure form. 
%**Perhaps state that appearance of c_{pl} alone comes from the Kirchhoff relation for L(T),
%as noted in Emanuel. 
}

{For condensation occurring at a grid point in the model without loss  of mass,  the term $pd\frac{1}{\rho}$ in Eq.(\ref{eq.enthalpy}) drops out, and there is a conserved quantity $\rho k - p$ before and after condensation. In a pure-steam atmosphere, this quantity (denoted by $\mathcal{A}$)    is  $\mathcal{A}\equiv\rho (c_{pl} T + Lq_c)- p$. We make use of this quantity to determine post-condensation temperature and pressure and the amount of condensates. } We start with a pre-condensation  situation wherein there is only gas with a temperature $T_1$ and   pressure $p_1$, and the gas is supersaturated due to radiative or adiabatic cooling. Then
\begin{equation}
    \mathcal{A} = \rho_1(c_{pl}T_1+L) - p_1 = \rho_1 c_{pl}T_1 + (\frac{L}{RT_1} - 1)p_1,
    \label{eq.a}
\end{equation}
where $\rho_1=\frac{p_1}{RT_1}$. After condensation, condensates form and temperature and pressure are adjusted, which are denoted as $T_2$   and $p_2$, respectively. At this point, the condensates are retained in the atmosphere. The temperature of the condensates is also $T_2$.
%as assumed when deriving the conserved quantity in Eq. (\ref{eq.enthalpy}).   
{$T_2$ and $p_2$ are restricted to the phase boundary $p_{sat}(T)$, given by the Clausius-Clayeyron relation,
because of the coexistence of condensed and vapour phases.}
%We don't need to state this -- later just say we used the constant-L form. 
%\begin{equation}
%    p_2 = p_{\ast}\exp{\left(-\frac{L}{R}\left(\frac{1}{T_2}-\frac{1}{T_{\ast}}\right)\right)},
%    \label{eq.cc}
%\end{equation}
%where $T_{\ast}$ is a reference temperature (372.76 K) and $p_{\ast}$ is a reference pressure ($10^5$ %Pa). Here we neglect the change of $L$ with respect to the change of temperature.
%because the typical temperature changes before and after condensation over one dynamical time step is %tiny. 
The conserved quantity in the post-condensation is then  
\begin{equation}
\begin{split}
    \mathcal{A} & \equiv \rho_2(c_{pl}T_2+Lq_c) - p_2 \\
    & = \rho_1 c_{pl}T_2+L\rho_c - p_2 \\
    & = \rho_1 c_{pl}T_2+\left(\frac{L}{RT_2}-1\right) p_{sat}(T_2),
    \label{eq.a2}
\end{split}
\end{equation}
in which we made use of conservation of mass $(\rho_1=\rho_2)$. With $\mathcal{A}$ given by the initial values using Eq. (\ref{eq.a}), we solve for $T_2$ in Eq. (\ref{eq.a2}) using Newton's method.
Because the range of temperatures encountered within the condensing layer is not large, we found it sufficient to use the analytic constant-$L$ form of $p_{sat}$, though it is an assumption that would be easy to relax. The condensate density $\rho_{\rm cond}$ is obtained from the difference of the gas density before and after the condensation.

With suitable cloud condensation nuclei, condensates form with a small supersaturation ratio $\mathcal{S}$ to overcome   surface tension \citep{houze2014},  in which $1+\mathcal{S}=p_1/p_{\rm sat}$. We allow $\mathcal{S}$ to be nonzero but still much smaller than 1. It is assumed to be $10^{-7}$ in most simulations to represent the  no-surface-barrier limit  but is varied up to 0.2 in some cases for sensitivity exploration. Note that the post-condensation pressure is adjusted to $p_{\rm sat}$.  

We do not model the detailed processes of cloud condensational growth and coagulation. Instead, the condensates are assumed to instantaneously  fall out after condensation occurs and evaporate in the subsaturated regions. { Evaporation is likewise treated using the conserved quantity $\mathcal{A}$.}  Pre-evaporation temperature and pressure in the evaporation regions are denoted as $T_3$ and $p_3$, and the total density is $\rho_3 = \frac{p_3}{RT_3}+\rho_{\rm cond}$, where $\rho_{\rm cond}$ is the density of the soon-to-be-evaporated precipitation. The conserved quantity is $\mathcal{A}=\rho_3(c_{pl}T_3+Lq_c)-p_3$, where $q_c=\frac{p_3}{RT_3}/\rho_3$. After evaporation, temperature $T_4$ and pressure $p_4$ are finally obtained via 
\begin{equation}
    \begin{split}
    \mathcal{A} & = \rho_3(c_{pl}T_4+L) - p_4 \\
    &   = \rho_3T_4(c_{pl}-R)+\rho_3 L.
    \label{eq.a3}
    \end{split}
\end{equation}
%The rate of evaporation should be sensitive to droplet size, relative humidity and temperature. 
We    adopt a highly idealized approach to determine the location of evaporation. All precipitation  evaporates within the atmosphere.  Evaporation   occurs only when the relative humidity is below 90\%. The evaporative mass  is assumed to evenly distribute over a fixed accumulative depth (this depth needs not  be continuous), and this depth is treated as a free parameter. Our fiducial models assume an evaporative depth of 10 km, but we will investigate the sensitivity of results to the evaporative depth.

There are two additional considerations to conserve energy associated with rainout. The first is  the dissipation of gravitational energy of the falling condensates. We assumed that this is dissipated via friction, and the frictional heating goes into the gas instantaneously. 
%The air temperature is adjusted using $T_{\rm new}\rho_{\rm air}c_v=T_{\rm old}\rho_{\rm air}c_v+g\rho_{\rm cond}\delta z$, where $\delta z$ is the model layer thickness, and the specific heat at constant volume $c_v$ is used because this is a constant volume process.  
The second is to consider the temperature difference between the condensates 
%(which has a temperature $T_2$ at the place where they form) 
and the air along the falling path. This heat exchange is assumed to occur via conduction and reaches equilibrium instantaneously. 
%The temperature is adjusted via $T_{\rm new}(\rho_{\rm air}c_v+\rho_{\rm cond}c_{pl})=(T_{\rm old}\rho_{\rm air}c_v+T_{\rm cond}\rho_{\rm cond}c_{pl})$. 
%This is a cooling effect on the atmosphere for   atmospheres that have a negative lapse rate $dT/dz$.

\subsection{Numerical setup and convergence}
We use the large-eddy-simulation setup that integrates the filtered Navier-Stokes equations. Stresses and fluxes on the resolved flow by  sub-grid turbulence are parameterized using a prognostic turbulence kinetic energy scheme similar to that used in \cite{deardorff1980}. Acoustic waves are treated explicitly in both horizontal and vertical directions using a time-splitting technique. The horizontal boundaries are periodic and the vertical boundaries are impermeable. A Rayleigh damping  is applied to winds in the top 10 km layers, and surface stress and heat flux are calculated using the original CM1 formulation at the bottom boundary. We applied a 6th-order numerical diffusion to maintain numerical stability. The model domain is in   Cartesian geometry with a typical grid space of 600 m in the horizontal  directions. The vertical grid space $\delta z$ ranges between 600 and 680 m depending on the surface temperature. Our canonical models have $320\times320\times250$ (in $x$, $y$ and $z$) grid points. The resolution is chosen to better resolve convection and waves while remaining computationally feasible.   Following the default values in CM1, we adopt the following constants: surface gravity $g=9.81\;{\rm m\;s^{-2}}$, the heat capacity at constant pressure  for water vapor $c_p = 1870 \jpkgk$ and for condensates $c_{pl} = 4190 \jpkgk$,  the specific gas constant  $R=461.9\jpkgk$, and latent heat $L=2.5\times10^6 \;{\rm J\;kg^{-1}}$.  The models are initialized  with a dry adiabat attached to the surface  and a moist adiabat when gas is saturated. Most simulations assume a 1-bar surface pressure which allows exploration of the essential dynamical features of the problem without the computational expense of a thicker atmosphere.  The setup corresponds to the behaviour of a planet which started with a low mass ocean; we can speculate that it would qualitatively mimic the behaviour of the upper bar of a much deeper atmosphere with a thicker noncondensing region.  { This setup treats the post-runaway climate that is interesting to a branch of 1D climate studies (e.g., \citealp{kasting1993,goldblatt2013,hamano2013,Boukrouche2021}), whereas the parametereization in \cite{ding2016} did not address this geometry.}
%The models are initialized  with a dry adiabat below the first condensation level with a bottom temperature 3 K cooler than the surface temperature, and a moist adiabat above the first condensation level. Small white-noise perturbations are added in the initial velocity field.

We ran a model with a   lower   resolution of 1 km in all directions    up to 88 simulation days. { The total kinetic energy reaches a statistical equilibrium after only a few model days. The total internal and potential energy reach a statistical equilibrium after about 25 days and then slightly oscillate around the mean.} However, quantitative properties of convective flows and gravity waves are insensitive to the long-term convergence of the model. For our canonical runs with higher resolution, we integrate the system only up to 5 to 11 simulation days as they are computationally more costly, and statistical results are obtained by outputs of the last 6 hours.

\begin{figure*}      % use "figure*" 
\epsscale{1.}      % adjust this number to change the size of your figure
\centering
\plotone{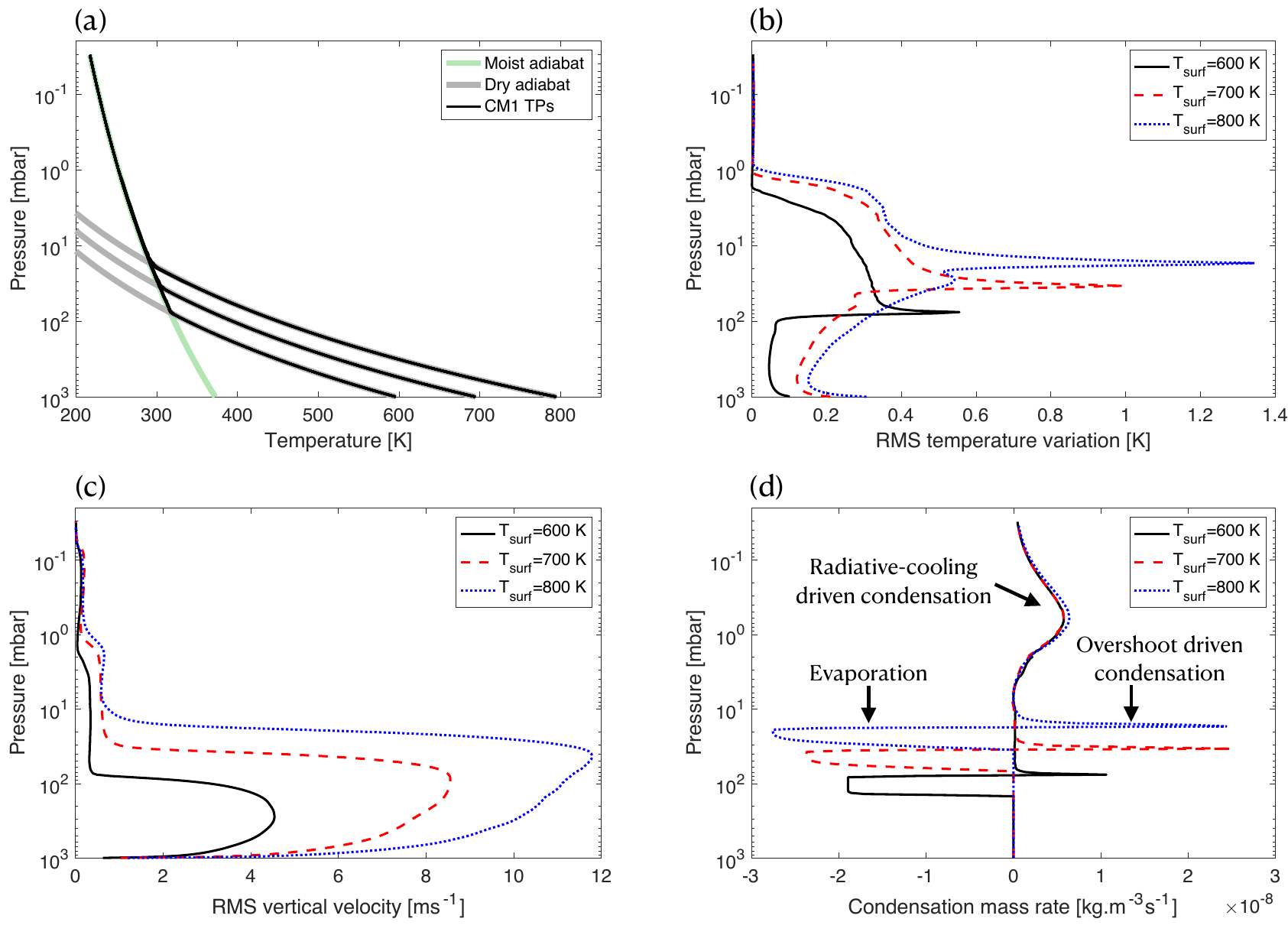}
\caption{Results from CM1 simulations with three surface temperature of 600 K, 700 K and 800 K, surface pressure of 1 bar and supersaturation ratio $\mathcal{S}=10^{-7}$, showing basic structure that is characterized by an upper condensing region and a lower dry convecting region. Statistical results are averaged over the last 6 hours of integration. \emph{Panel (a):} randomly selected instantaneous temperature-pressure profiles (black lines). Note that there are multiple black lines per case but they have very small horizontal temperature differences and so they appear to be merged. Thick grey lines are dry adiabats corresponding to different surface temperatures; the thick green line is the moist adiabat for pure-water atmosphere. \emph{Panel (b):} horizontal- and time-averaged root mean square (RMS)  temperature variations as a function of mean pressure. \emph{Panel (c):} horizontal- and time-averaged RMS vertical velocities as a function of mean pressure.  \emph{Panel (d):} horizontal- and time-mean condensation mass rate (positive meaning condensation and negative meaning evaporation) as a function of mean pressure. }
\label{fig.1ds0}
\end{figure*} 

\section{Results}
\label{ch.results}

\subsection{Basic structure and dynamics}
\label{ch.basic}

Our simulations show that the atmospheric domain is generally comprised of two characteristic regions --- a relatively quiescent, stratified upper condensing region and a vigorously convecting lower dry region, separated by the first condensation level. This is similar to those found in 2D convection simulations of \cite{yamashita2016}. We start with describing horizontal-mean properties of the simulations. Figure \ref{fig.1ds0} shows results as a function of mean pressure   from three models with   a surface pressure of 1 bar, supersaturation ratio $\mathcal{S}=10^{-7}$ and  three surface temperatures of 600, 700 and 800 K. In both condensing and dry regions, the   horizontal root-mean-square (RMS) temperature variations are typically much smaller than 1 K (panel b) in the domain except near the first condensation level where  convective overshoots occur. The instantaneous temperature-pressure profiles in  panel (a) of Figure \ref{fig.1ds0} appear to be merged to a single dry adiabat in the lower region and to the moist adiabat in the upper region. The RMS vertical velocities   in panel (c)  reach several to more than 10 $\mps$ in the lower dry region  but rapidly decrease above the first condensation level. The overall vertical speed increases with increasing surface temperature.  The domain-mean  condensation/evaporation mass rates  in  panel (d) show  evaporation below the first condensation level but a sharp transition to condensation right above the first condensation level.  The latter is driven by the convective overshoots and  gravity waves. Low pressures ($\lesssim 1$ mbar) show a vertically broad and smooth condensing region that is primarily driven by radiative cooling.

The upper condensing zone further exhibits two subdivisions. One is above the first condensation level but below the radiative cooling zone. This region is  optically thick and has small radiative cooling  rates (on the order of $10^{-5}\;{\rm Ks^{-1}}$). While the ascending branch  of wave motions  supersaturates and is adjusted to saturation, the descending branch is free of thermal damping.  Waves can vertically propagate through this  region and exert certain vertical velocity and temperature fluctuations. The other subdivision at the thermal photosphere ($\lesssim 1$ mbar) show negligible RMS temperature and vertical velocity. Radiatively driven condensation acts uniformly in the horizontal direction and sets stringent constraints on the local condition.
%in the presence of a negligible supersaturation ratio $\mathcal{S}$. 
The collapse of pure-steam gas  into a unique property determined by the   Clausius-Clapeyron relation eliminates dynamical perturbations \citep{pierrehumbert2016}.

\begin{figure*}
\epsscale{1.}
\centering
\plotone{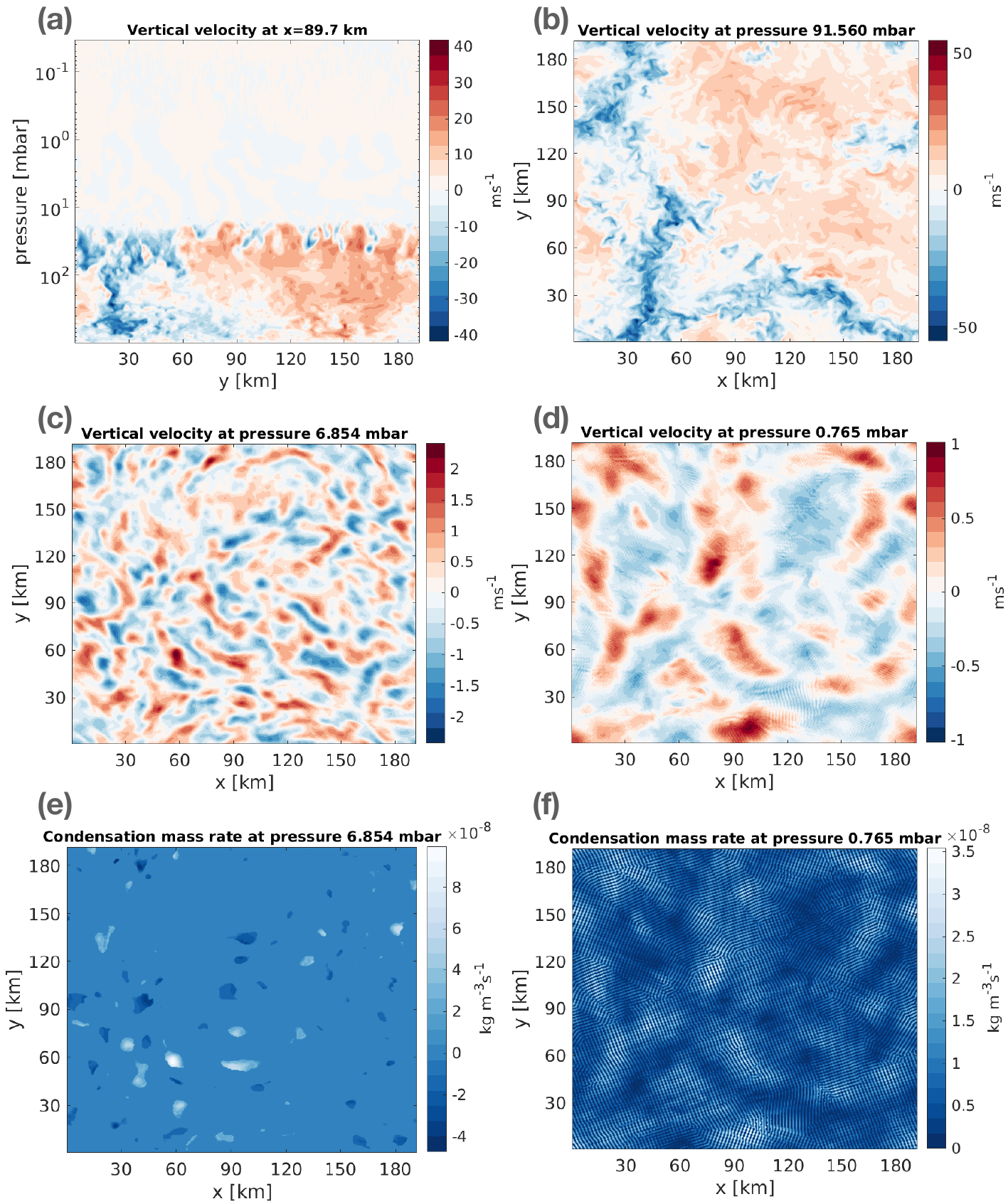}
\caption{Instantaneous snapshots of the model with  surface temperature of 800 K, surface pressure of 1 bar and  supersaturation ratio $\mathcal{S}=10^{-7}$. \emph{Panel (a):} Vertical velocity at a slice as a function of $y$ and pressure. \emph{Panel (b) to (d)}: Vertical velocities at different pressure levels, with (b) representing those at the lower dry convective zone, (c) at the overshooting zone slightly above the first condensation level, and (d) at the radiatively cooling zone. \emph{Panel (e) and (f):} condensation mass rate at different pressure levels with (e) in the overshooting zone and (f) in the radiative cooling zone. }
\label{fig.maps_t800s0}
\end{figure*}

{ Note that since the condensation scheme is applied at each time step, and condensate is not allowed to accumulate, at most a small portion of the vapour is converted to condensate and subsequently removed by precipitation. There is never a large proportion of atmospheric mass removed (see the small condensation mass rate in panel (d) of Figure \ref{fig.1ds0}); the interesting aspect of diluteness is that the small mass loss at each condensation step can add up over time to transport a significant proportion of atmospheric mass. The condensed phase occupies very little volume, so its removal does not significantly change the $(p,T)$ of the gas phase.  In a non-hydrostatic model, removal of precipitation pushes the atmospheric column  out of hydrostatic balance, but the dynamics handles the resulting pressure adjustment
explicitly. If the expansion caused by unburdening lower air parcels causes cooling and supersaturation, that will be handled by condensation at subsequent time steps.}

Now we present the 3D  structure of convection and condensation. Dynamics in the lower noncondensing  region is characterized by a major convective overturning cell that spans across the whole horizontal domain and with peak velocities reaching several tens of $\mps$. Negative buoyancy is generated in the upper parts of the noncondensing region by evaporative cooling, resulting in denser currents that penetrate deep down. 
Figure \ref{fig.maps_t800s0} displaces snapshots of vertical velocity in a vertical cross section and several horizontal cross sections, as well as condensation mass rates in two cross sections from the model with a surface temperature of 800 K.  Panel (a) and (b) show a broad and coherent  upwelling cell and narrower ridges of downwelling. As a consequence of the area   asymmetry and continuity, the downdrafts have larger speeds than the updrafts.  

{ In the condensing region, there is no deep penetrating plume even though latent heat is released, which is in stark contrast to the deep moist convection on Earth. The lack  of buoyancy generation in the pure-steam atmosphere is  because once the atmosphere is saturated, the  temperature-pressure profiles collapse into a single adiabat that is uniquely determined by the Clausius–Clapeyron relation.  The condensing air parcel is neutrally stable with regard to the environment because the air parcel shares the same density-pressure trajectory as the environment  \citep{colaprete2003,ding2016}. On the other hand, the condensing region is strongly stratified against downward motions that are not associated with condensation. While the  saturated air may still have the possibility to move freely upward as it is neutrally buoyant, its motions will be limited by the stratification experienced by downward-moving air via mass continuity.  This excludes convective instability in the saturated region even with vigorous perturbations from  the lower noncondensing layer. Behaviour of this sort was hypothesized in the 1D treatment of \citep{ding2016}, but the nonhydrostatic model is free to deviate from it; the confirmation of the supposition in a fully dynamic model is thus significant.  }

In our simulations all the condensation and cloud formation is generated by small vertical displacements by gravity waves or by radiatively driven condensation in the thermal photosphere.    Internal gravity waves are triggered by the dry convective penetration and show much smaller spatial structure and velocity magnitudes than those in the dry convective region.  Panel (c) and (d) show characteristic wave patterns. Gravity waves generated from convective perturbation have long been identified in Earth and planetary atmospheric modeling (e.g., \citealp{fovell1992, baker2000,lefevre2018}). Time evolution of these fields show  a good correlation between locations of upwelling convective plume and sources of gravity waves. Gravity-wave induced temperature variations can either trigger condensation in the ascending regions  or permit evaporation in the descending regions.  Panel (e) illustrates that precipitation fall from higher altitudes can evaporate at locations with downward velocity (and thereby are hotter and subsaturated), whereas regions with upward velocity can generate condensation. In the radiative zone shown in panel (f),  only condensation can occur and it is more widespread. Its spatial pattern generally correlates well with the gravity waves.

\subsection{Sensitivity exploration}
We perform experiments with varying parameters to evaluate the qualitative and quantitative differences of models to those shown in Section \ref{ch.basic}. These experiments have the same parameters and setup as the one with a surface pressure of 1 bar, a surface temperature of 800 K, a supersaturation ratio $\mathcal{S}=10^{-7}$, evaporation depth of 10 km and no rotation, except those specified in the following text. We show that the basic structure and dynamics remain qualitatively the same, but the quantitative dynamical details can differ in some cases. 

\begin{figure*}
\epsscale{1.1}
\centering
\plotone{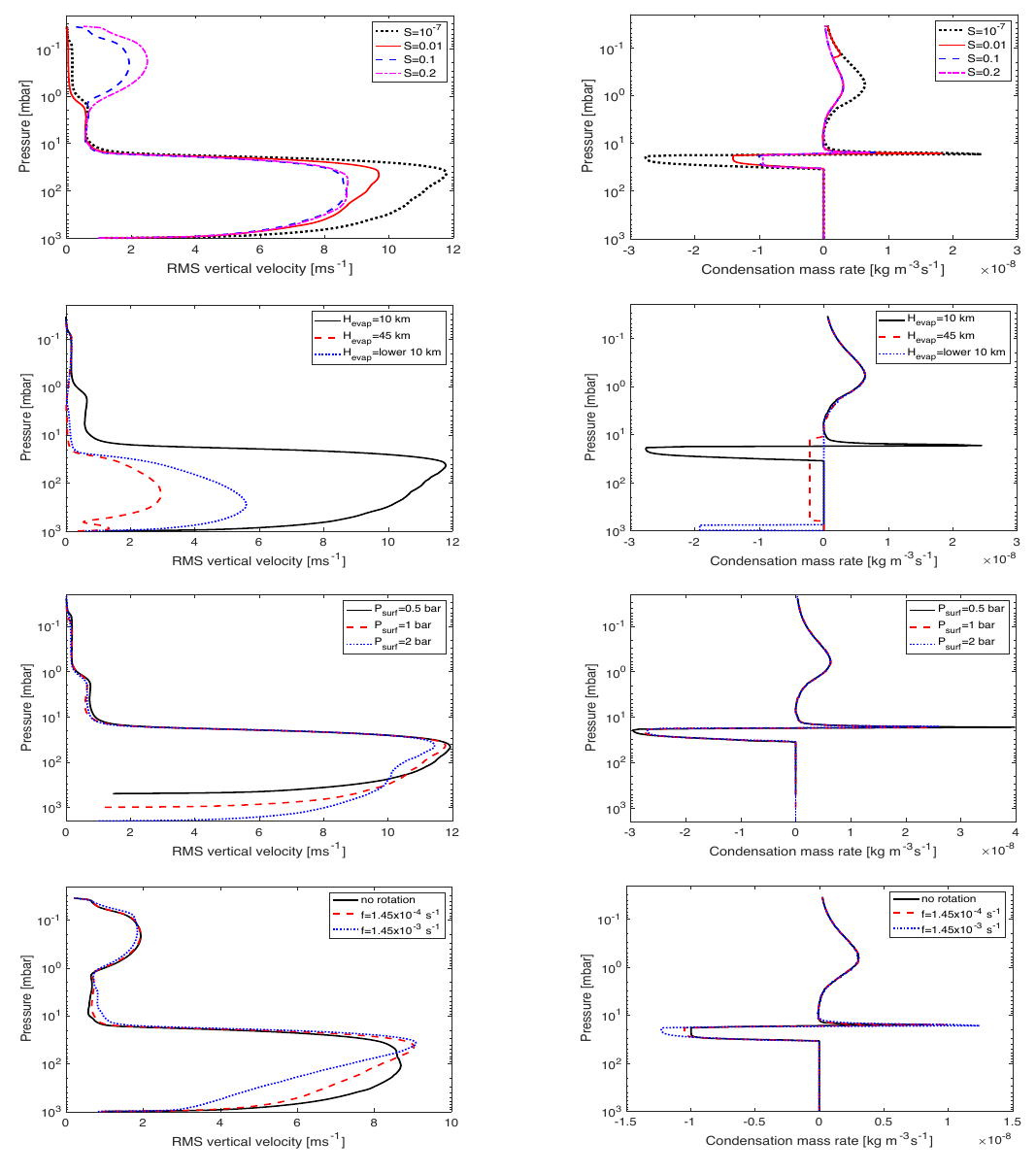}
\caption{
Horizontal- and time-averaged RMS vertical velocity (left column) and condensation mass rate (right column) as a function of pressure for various models. \emph{First row:} models with surface temperature of 800 K and surface pressure of 1 bar, but with four supersaturation ratio $\mathcal{S}=10^{-7},~0.01, ~0.1$ and 0.2.  \emph{Second row:} models with different evaporation height and location indicated by the thicknesses and locations of the evaporative mass rates on the right. \emph{Third row:} models with different surface pressures but with the same dry adiabat as that with a surface temperature of 800 K and surface pressure of 1 bar. \emph{Bottom row:} models with zero and two different Coriolis parameter $f=1.45\times10^{-4}$ and $1.45\times10^{-3} \;{\rm s^{-1}}$, and a supersaturation ratio $\mathcal{S}=0.1$.
}
\label{fig.w_condmass_all}
\end{figure*}

We first conduct additional experiments with different supersaturation ratio  $\mathcal{S}=0.01$, 0.1 and 0.2.   In the pure-steam condition, the characteristic permitted   temperature variation before and after condensation is $\delta T\approx \mathcal{S}T^2R/L$ when $\mathcal{S}\ll 1$. Higher $\mathcal{S}$  permits    more flexibility of  temperature variation in the condensing region, and potentially stronger waves there. The first row in Figure \ref{fig.w_condmass_all} shows results of the experiments. Indeed, in the upper radiative cooling zone ($\lesssim 1$ mbar), higher $\mathcal{S}$ leads to larger RMS vertical velocity (left panel) and larger RMS temperature variation ($\gtrsim0.2$ K for $\mathcal{S}\geq0.1$, not shown). 
%Note that, the temperature profiles in these experiments still collapse to a moist adiabat in the condensing region like that shown in panel (a) of Figure \ref{fig.1ds0}.  
Somewhat surprisingly, the RMS vertical velocity at the region below the radiative cooling zone and above the dry convective zone remain invariant with different $\mathcal{S}$. In the lower dry region, the RMS vertical velocity slightly decreases with increasing $\mathcal{S}$, although the decrement appears to saturate when $\mathcal{S}\gtrapprox0.1$. 
This results in weaker overshoot-driven condensation when  $\mathcal{S}$ is larger as shown on the right.

At a given time, condensation in the radiative cooling region is widespread in the case with $\mathcal{S}=10^{-7}$ but is highly sparse when $\mathcal{S}$ is no longer tiny (not shown). Dynamically, when a certain region condenses and  adjusts back to a saturated state, its perturbation  rapidly propagates. This maintains other regions  ``warm" such that condensation does not easily occur.  Energetically, heat released from a single condensing event  increases with increasing $\mathcal{S}$. 
%If condensational heating tends to balance radiative cooling, it is natural that 
Given the same radiative cooling rates, area fraction of condensation decreases with increasing $\mathcal{S}$ to balance radiative cooling. Interestingly,  the case  with $\mathcal{S}=10^{-7}$ shows a larger domain-mean condensation rate at the photosphere that is almost balanced by radiative cooling, while others   with $\mathcal{S}\gtrapprox 0.01$ show  smaller condensation rates. %The former implies that condensation and precipitation are a major avenue of energy transport in response to the radiative cooling, 
This implies that gas dynamics becomes increasingly important in energy transport  with a small but non-negligible supersaturation ratio. %In the latter case, still, gravity waves, rather than convectively unstable motions, are the dominant dynamics in the upper condensing region.

Next, we present experiments with varying  location and depth of the evaporation shown in the second row of Figure \ref{fig.w_condmass_all}. These affect the the generation of negative buoyancy in the dry convective region because evaporation is the major cooling mechanism to drive the dry convection. In the first experiment, we extend the evaporation depth to 45 km as appose to the original 10 km; and in the second case, evaporation is only permitted  in the lower 10 km   above the ground. These settings can be visualized by the evaporative mass rate profiles shown on the right column.   In both cases, although the basic structure and characteristic dynamics remain the same, the magnitude of the velocities is smaller than the canonical case. This is not surprising as negative buoyancy are expected to be smaller in both cases.  The case with an evaporation depth of 45 km shows the weakest activity, with RMS vertical velocity   $<3\mps$. The case with a lower 10 km evaporation depth is in between. The RMS vertical velocities above the first condensation level of both experiments are negligible compared to the nominal case, which is related to the weaker penetrative plume at the top of the dry convective zone and therefore a weaker generation of gravity waves. As a result, both cases show little overshoot-driven condensation. In terms of the spatial organization,  the dry convective cell spans the whole horizontal domain in all cases (not shown). But the updrafts and downdrafts in the case with an evaporation depth of 45 km appear  to be more symmetric than the other two cases.

We then carry out experiments with a varying surface pressure of 0.5 bar and 2 bars adjusting the surface temperature so as to keep  the same dry adiabatic profile  as that of the nominal case. The similarity of both the RMS vertical velocities and condensation mass rates shown in the third row of Figure \ref{fig.w_condmass_all} demonstrates that this level of variation on the surface pressure changes  the quantitative dynamics little. This is a bit surprising as we would have expected that the higher the surface pressure, the more vigorous the dry convection   because  more vertical length would facilitate larger convective available potential energy for a nondiluted convective plume. Perhaps the dry convection is sufficiently turbulent  (see Figure \ref{fig.maps_t800s0}) that such  undiluted convective plumes are not present.  This experiment also suggests that the behavior near and in the condensing layer is not too sensitive to processes occurring near the surface or in the  noncondensing deeper layers.

Lastly, we  examine the role of rotation in shaping the convection and whether  coherent vortex would form in  small-scale nonhydrostatic models in pure-steam conditions. One might first expect vigorous hurricanes to form because the atmosphere already contains enormous quantities of latent heat, whereas for Earth-like hurricanes the latent heat has to be imported by evaporation from the ocean surface \citep{emanuel2018}.  3D radiative-convective equilibrium modeling applied in Earth's tropical conditions suggest that rotation triggers hurricane-like vortices (e.g., \citealp{bretherton2005,held2008,merlis2016}). However, GCMs with rich condensable vapor and  Earth's  rotation rate did not generate hurricane-like features \citep{pierrehumbert2016}. It is unclear whether that is due to the hydrostatic nature of GCMs, or low horizontal resolution, or  an intrinsic dynamical property of condensable-rich atmospheres. Here, we explore two Coriolis parameters,  $f=1.45\times10^{-4}\;{\rm s^{-1}}$ (the value at Earth's North pole) and another one 10 times of that $f=1.45\times10^{-3}\;{\rm s^{-1}}$.  The domain-mean properties are shown in the bottom row of Figure \ref{fig.w_condmass_all}. All three models have a supersaturation ratio $\mathcal{S}=0.1$ because we expect that models with higher flexibility of temperature have a higher chance of being sculpted by the rotation. The inclusion of rotation results in small changes in the   RMS vertical velocity and condensation rate, and a slight change in the vertical profile of the RMS vertical velocity in the dry convective zone. Similarly, differences in the condensation mass rates are small.  

\begin{figure}
\epsscale{1.1}
\centering
\plotone{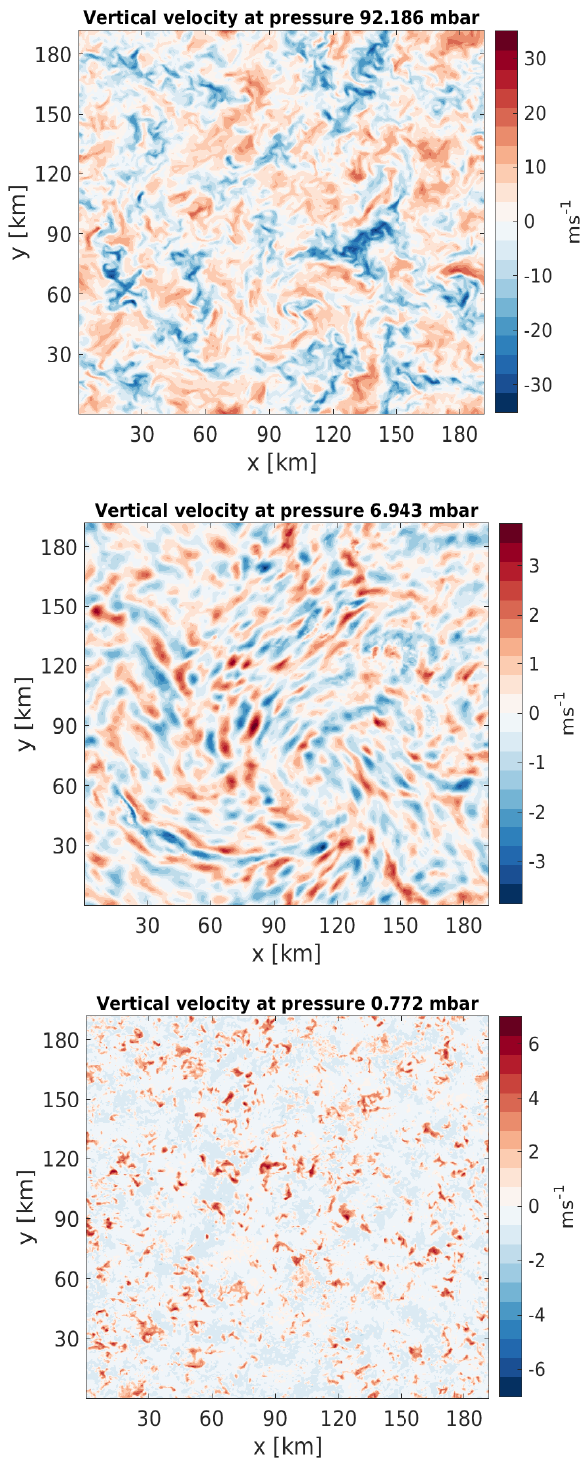}
\caption{Snapshots of vertical velocities at   the dry convective zone (upper panel), the overshooting zone (middle panel) and the radiative cooling zone (bottom panel) from the model with surface temperature 800 K, surface pressure of 1 bar and supersaturation ratio $\mathcal{S}=0.1$, and including rotation with $f=1.45\times10^{-3}\;{\rm s^{-1}}$. Pressures are indicated above each panel. }
\label{fig.f}
\end{figure}

We do not find existence of long-lasting, coherent vortices.  Figure \ref{fig.f} shows snapshots of vertical velocity at different pressure levels from the model with $f=1.45\times10^{-3}\;{\rm s^{-1}}$.  The strong rotation only tends to limit the horizontal size of dry convective structures. 
%The Rossby number $Ro = U/(fL)$ in this case is small. Adopting a convective speed $U\sim 10\mps$ and a domain length scale $L\sim 2\times10^5$ m, $Ro\sim0.3$. This implies that the Coriolis force is important in the force balance.
Rotation could result in swirling structures in the dry convection (for instance, the feature at around x=100 km and y=50 km in the upper panel). Wave properties in the upper condensing zone are also  influenced by rotation. Perhaps the lack of vortex formation is not surprising because a systematic baroclinic  dynamical structure is likely needed to maintain the available potential energy against dissipation due to the surface drag and related vortex dissipation mechanism (such as the Ekman pumping). Such a baroclinic structure  is difficult to maintain in the pure-steam atmospheres due to the strong constraint of the Clausius-Clapeyron relation \citep{colaprete2003,ding2016,pierrehumbert2016}.

\section{Discussion}
\label{ch.discussion}
The lack of convective instability and organized convective structure in the condensing region of pure-steam atmospheres suggests that convection parameterization in both relevant GCMs and 1D models may safely neglect the role of deep penetrating plumes and mainly consider condensation, precipitation and evaporation. The basic parameterization scheme proposed by \cite{ding2016} and its application in \cite{pierrehumbert2016} should be sufficient for atmospheres closed to the pure-steam limit. The domain-mean temperature structure of our full 3D simulations maintains the dry and moist adiabats as those typically assumed for 1D models, which is encouraging for 1D radiative-convective models for post-runaway climate calculations using sophisticated radiative transfer (e.g., \citealp{goldblatt2013,kopparapu2013, Boukrouche2021}).
{ GCM simulations of steam atmospheres in the  early stages of Earth and Venus before water vapor has condensed into an ocean  illustrates the importance of understanding climate dynamics of such atmospheres \cite{turbet2021}. Our configuration is similar to that in these GCM calculations  except for the lack of the suppression of condensation and dry convection by shortwave heating in the intermediate layer. This intermediate stratified layer  serves to suppress the strong convective perturbations on the bottom of the saturated layer and  eliminates additional thick cloud formation near there due to these overshooting, which supports results in \cite{turbet2021}. }

Although our models do not include cloud radiative forcing, our results provide implications for cloud parameterization and observatonal implications for atmospheres closed to the pure-steam limit. Condensation is driven by radiative cooling in the thermal photosphere, though some does happen deeper, near the boundary of the condensing region, due to triggering by gravity waves. In the photosphere, radiative cooling  acts  uniformly in the horizontal direction, which implies that clouds tend to be statistically homogenized at the photosphere of steam atmospheres. We do not see evidence for aggregation of condensing patches in the thermal photosphere. The lower condensing region below the photosphere is characterized by convective penetration and gravity waves. The resulting   cloud structure would be spatially correlated to the convective plumes and waves. Although these cloud structures cloud would be highly time variable and patchy, the horizontal mean cloud mixing ratio can be comparable to those driven by radiative cooling (see the condensation rate in panel d of Figure \ref{fig.1ds0}).   Such possible cloud configuration ---most clouds in a steam atmosphere would form in the thermal photosphere, though they can nonetheless exhibit random patchiness (see Figure \ref{fig.maps_t800s0} d and f)--- should have observational implications for observations of steam atmospheres of both sub Neptunes and terrestrial exoplanets in the runaway stage. Of course,  the amount of small cloud droplets that are able to keep aloft is sensitive to microphysics, which in turn is affected by the mode of vertical transport. This aspect can be quantified only with  fully coupled dynamics-microphysics models. Finally, the suppression of deep convection in the condensing layer implies weak transport of trace species from the deep atmosphere, which would have interesting implications for the chemistry of such atmospheres.

%We have tried to diagnose mass and energy transport using the a fully equilibrated model but with lower surface pressure. We expect that domain-mean slow upwelling motions balance the mass redistribution due to precipitation and evaporation. However, our diagnoses do not show this. There are two reasons. First, mass lost due to numerical integration and the mass regainment imposed by CM1 represent an additional numerical mass redistribution. Second, and perhaps more critically, acoustic waves are integrated separately using the smaller timesteps, meaning that outputs only record motions that are non-acoustic. The significant mismatch between our diagnoses and prediction indicates that acoustic waves is an important mechanism to adjust mass redistributed by precipitaton and evaporation.

To summarize, in this study, we have performed high-resolution,  non-hydrostatic  simulations to simulate convection in pure-steam atmospheres. We find that the atmosphere is characterized by an upper condensing region and a lower dry convecting region. The condensing region is stratified and characterized by gravity waves that are triggered by convective overshoots.  The lower dry region is characterized by convective overturning cells that span  over the model domain.  Magnitude  of  velocities in the condensing region is much smaller than that in the lower dry region. Horizontal temperature variation is small ($\lesssim1$ K) overall. Condensation in the thermal photosphere  is driven by radiative cooling and tends to be statistically homogeneous. Near the boundary of the condensing region, condensation is triggered by gravity waves and convective penetrations and  exhibit random patchiness. Models with different parameters show a similar qualitative picture but can be quantitatively different in some cases.   Our results should also be applicable in exoplanets with thick gaseous envelopes filled with mostly condensable species.

Future extensions would be to include various fractions of non-condensable gases and to explore convection from dilute to non-dilute conditions with potentially deeper atmospheres. Effects of clouds, including mass loading and radiative effects in both visible and thermal bands, may yield more complexity and feedbacks in controlling the convective systems. { Realistic cloud microphysics schemes might yield a time delay of the full relaxation back to the phase equilibrium. This could affect the amount of condensates in the air as well as the heating or cooling rates and therefore influence the generation of buoyancy on the convective system.  }
If water vapor absorption of instellation is included, it would reduce the net radiative cooling rate in the condensing layer, and the convective mass transport in the subsaturated noncondensing layer (e.g., \citealp{turbet2021}). This aspect would be especially interesting for planets around M dwarfs whose peak spectral energy is close to near-IR.
Gravity waves in our current model are obtained without large-scale vertical wind shear. Its existence could impact the generation and vertical propagation of the waves as well as cloud formation \citep{lefevre2020}, which is worth examining in future studies. Finally, the suppression of convection in the condensing layer implies weak transport of trace species from the deep atmosphere, which may have interesting implications for the chemistry of such atmospheres. 

\acknowledgements
%X.T., M.L. and R.T.P. are  supported by  the European community through the ERC advanced grant EXOCONDENSE (PI: R.T. Pierrehumbert). 
This project has received funding from the European Research Council (ERC) under the European Union’s Horizon 2020 research and innovation program (grant agreement No. 740963/EXOCONDENSE).

\if\bibinc n
\bibliography{draft}
\fi

\if\bibinc y

\fi

\end{document}